\documentclass[a4paper,fleqn,usenatbib]{mnras}

\usepackage[T1]{fontenc}
\usepackage{ae,aecompl} 

\usepackage{graphicx,epstopdf}   
\usepackage{amsmath}    
\usepackage{amssymb}    
\newcommand\tab[1][1cm]{\hspace*{#1}}

\title[Network Analysis]{Network analysis of the COSMOS galaxy field}

\author[R.~de Regt et al]{ R.~de Regt$^{1,2}$,\thanks{E-mail:{deregtr@uni.coventry.ac.uk}}
  S.~Apunevych$^{3}$,
  C.~von Ferber$^{1,2}$,
  Yu.~Holovatch$^{4,2}$,
  B.~Novosyadlyj$^{3,5}$
\\
$^{1}$Applied Mathematics Research Centre, Coventry University,
Coventry, CV1 5FB, UK\\
$^{2}$ $\mathbb{L}^4$ Collaboration \& Doctoral College for the Statistical
Physics of Complex Systems, Leipzig-Lorraine-Lviv-Coventry,
\\ \hspace{0.75em} D-04009  Leipzig, Germany \\
$^{3}$Ivan Franko National University of
Lviv, Kyryla i Methodia Str., 8, UA-79005 Lviv,
Ukraine\\
$^{4}$Institute for Condensed Matter Physics, National
Acad. Sci. of Ukraine,UA-79011 Lviv, Ukraine\\
$^5$International Center of Future Science of Jilin University, Qianjin Street 2699, Changchun, 130012, P.R.China}
\date{Version: 2, Accepted 2018 March 18. Received 2018 February 24; in original form 2017 July 3}

\pubyear{2018}

\begin{document}
\label{firstpage}
\pagerange{\pageref{firstpage}--\pageref{lastpage}}
\maketitle

\begin{abstract}

  The galaxy data provided by COSMOS survey for
  $1{^\circ} \times 1{^\circ}$ field of sky are analysed by
  methods of complex networks.  Three galaxy samples (slices) with
  redshifts ranging within intervals 0.88$\div$0.91, 0.91$\div$0.94
  and 0.94$\div$0.97 are studied as two-dimensional projections for the
  spatial distributions of galaxies. We construct networks and
  calculate network measures for each sample, in
  order to analyse the network similarity of different
  samples, distinguish various topological environments, 
  and find associations between galaxy properties (colour
  index and stellar mass) and their topological environments.

  Results indicate a high level of similarity
  between geometry and topology for different galaxy samples and no
  clear evidence of evolutionary trends in network measures. 
  The distribution of local clustering coefficient $C$
  manifests three modes which allow for discrimination between
  stand-alone singlets and dumbbells ($0\leq C \leq 0.1$),
  intermediately packed ($0.1 < C < 0.9$) and clique
  ($0.9 \leq C \leq 1$) like galaxies.

  Analysing astrophysical properties of galaxies (colour
  index and stellar masses), we show that distributions are
  similar in all slices, however weak evolutionary trends can also be
  seen across redshift slices. To specify different topological
  environments we have extracted selections of galaxies from each sample
  according to different modes of $C$ distribution. We have found statistically
  significant associations between evolutionary parameters of galaxies
  and selections of $C$: the distribution of stellar mass for galaxies with
  interim $C$ differ from the corresponding distributions for
  stand-alone and clique galaxies, and this difference holds for all
  redshift slices. The colour index realises somewhat different behaviour.
\end{abstract}
\begin{keywords}
  cosmology, large-scale structure of the Universe, galaxy
  evolution 
\end{keywords}

\section{Introduction}

The observable large-scale structure of the Universe appears to be rich in a variety of shapes and
topological features, we can identify clusters, super clusters, voids,
walls and filaments in it. Altogether they comprise the Cosmic Web,
the term coined in \cite{Bond96}.  Numerous approaches have been devised in
an attempt to properly describe and analyse the geometry and topology
of the Cosmic Web, see for example the recent studies
\citep{Cautun14, Chen15, Chen16, Hahn14, Leclercq16, Lee16, Pace15,
  Pranav16, Ramachandra16, Zhao15, Libeskind17}. \\
\indent Methods and approaches of network science, see \citep{Albert2002,
  Dorogovtsev2003, Barat2008, Newman2010, rt}, have
recently proliferated into various disciplines including cosmology,
\citep{Hong15, Hong2016, Coutinho2016}. Complex networks are believed
to assist in solving various open problems of cosmology, e.g. clarify
the impact of environment on galaxy evolution
\citep{Brouwer16,Kuutma17}; quantify the geometry and topology of
large scale structures; understand the formation of phase-space
distribution of dark and luminous matter and thus reveal the nature
and properties of dark matter and dark energy.
\\
\indent
The aim of the present paper is to study of the observable Cosmic Web with the
aid of complex networks, develop and validate a universal approach
for extracting topological environments from the observational data, in
order to investigate the relation between properties of a galaxy and
its place in large-scale structures, such as clusters, voids, walls 
etc.

We follow the pioneering paper by \cite{Hong15} where three network
measures of topological importance (degree centrality, closeness
centrality and betweenness centrality) have been derived for one
galaxy sample from the COSMOS catalogue \cite{Ilbert13}, different
topological environments in the Cosmic Web have been selected and their
relationship to evolutionary parameters has been estimated.  This
paper \cite{Hong15} in turn follows \cite{Scoville13}, where the
same problem was addressed using ``traditional'' methods and the same
data source.

In comparison with already existing methods developed for Cosmic Web analysis, network analysis has a number of potential benefits: (a) it is not built on some ad-hoc assumptions on the nature of the data, e.g. existence of a continuous density field; (b) it's computationally effective in treating discrete data, as no density estimator or Hessian is computed; (c) it is capable of describing and quantifying the content of data at an adjustable level of detail and complexity, properly encoding information; (d) it's equally applicable to results of simulations and real observational data, thus allowing for direct comparison between them; (e) it can go beyond the classification of environments as clusters or filaments, by providing a more holistic view on the topology of the multi-scale phenomenon of the Cosmic Web. Thus, network analysis can complement other methods and effectively integrate them into a framework capable of investigating the complexity of large-scale structures of the Universe.

Here, we extend network analysis to include several galaxy samples and
compare constructed networks by introducing other network metrics
of interest like: number of edges, mean node degree, size of the
giant connected component, average path length and diameter,
assortativity. Also, we advocate the usage of clustering coefficient
as a measure of short-range order which provides a robust 
technique that can be applied to generate networks for real observational
data and simulation outputs. Moreover, we assess the applicability, restrictions and accuracy of such a technique.

The paper is organised as follows. In Section 2 we describe the
observational data, the methodology of network construction and
analysis is summarised in Section 3. Section 4 is devoted to results
and discussion. The conclusions to be found in Section 5.

\section{COSMOS samples of galaxies}

The COSMOS Collaboration\footnote{http://cosmos.astro.caltech.edu} is
a grand astronomical endeavour which seeks to integrate data produced
by a variety of space and ground-based telescopes. The survey is aimed
at analysing galaxy evolution and designed to collect essentially all
possible objects in the field of view, i.e. to be as deep as possible,
meanwhile covering an area of celestial sphere large enough to
mitigate for the influence of cosmic variance.

The datasets for exploration are driven from the catalogue built by
\cite{Ilbert13} on the base of UltraVISTA ultra-deep near-infrared
survey, data release DR1 \cite{McCracken12}. It includes directly
observable quantities, such as celestial coordinates for galaxies and
photometric magnitudes for a number of broad bands, as well as colour
corrected for dust extinction, $M_{NUV} - M_{R}$. Moreover, the dataset
includes indirect estimations obtained by fitting models to
photometric data \cite{Ilbert13}: most important is $z$, the redshifts
for galaxies; basic galaxy classification according to colour --
quiescent or star-forming; and other physical parameters of galaxies,
e.g. stellar mass.

This catalogue was built for studying the mass assembly of galaxies
\cite{Ilbert13}, used for exploring the evolution of galaxies and
their environments in \cite{Scoville13}, as well as for constructing
complex networks \cite{Hong15}. Thus, this data set could be
considered as a standard for benchmarking different kinds of
large-scale structure analyses.

To achieve the goals of this study, we require independent samples 
of galaxies, meaning each sample should contain a
unique set of galaxies. The samples should also approximately represent the same statistical population, to ensure comparisons statistically viable. As the survey covers quite a modest area of celestial sphere, the optimal region to be chosen lies in the centre of surveyed area where the right ascension (R.A.) spans the range
$149\overset{^\circ}{.}\!4\div159\overset{^\circ}{.}\!4$ and declination (Decl.) is in the range $1\overset{^\circ}{.}\!7\div2\overset{^\circ}{.}\!7$.

The samples of galaxies are derived from the data set considering neighbouring ranges of redshift: $0.88\le z<0.91$, $0.91 \le z \le 0.94$ and $0.94< z\le0.97$, to be referred hereafter as $z_1$, $z_2$ and $z_3$ respectively. By this choice we extend the data analysed by \cite{Hong15} for redshift $z_2$ to include neighbouring
redshift slices $z_1$ and $z_3$. Such an extension should minimise the
influence of selection effects meanwhile providing large enough populations of different types of galaxies, including a high proportion of early-type
(red) galaxies. Also, the central slice reproduces the one used in
\cite{Scoville13}, where it was shown that when $z>1$ the relation of
galaxy properties within a local environment abruptly diminishes.

The elaborated analysis of multi-band photometry data estimates the
redshifts of galaxies to a high degree of accuracy (at 1\% level). The
thickness of slicing (redshift intervals $\Delta z$) is chosen to be
comparable to the errors in $z$ and to ensure a large enough sample of galaxies to make statistical methods meaningful.

For the standard $\Lambda$CDM cosmology with $H_0$=70 km/s/Mpc and
$\Omega_{\Lambda}=0.7$, a one degree distance on celestial sphere at
$z=0.91$ corresponds to a distance of $\approx54$ Mpc. Whilst 
the redshift interval $\Delta z=0.03$ corresponds to a spatial
thickness of $\approx76$ Mpc in comoving spatial coordinates.
Despite the progress made in redshift determination, its accuracy is still
insufficient to allow for three-dimensional spatial analysis. Thus, we analyse the each redshift as a two-dimensional projection of celestial sphere. This projection brings about some additional systematic bias and noise distorting the cosmic network. A more detailed discussion of such effects for density estimations can be found in
\cite{Scoville13}.

Given the above mentioned restrictions of the data set, we still believe the data is good enough for answering the major questions at hand and
validating the approach. The forthcoming releases of COSMOS and other
extragalactic surveys can potentially mitigate or even remove such
restrictions.

\section{Methods of network analysis}

\subsection{Network construction}
\label{sect:NetworkConstruction}

Contrary to the data coming from computer science, industrial
databases and social networks, the data in cosmology are inherently
non-networked and contains a substantial amount of noise. Hence, a
graph (network) must be constructed from the data set (catalogue)
using appropriate criteria and methodology, and preferably without
losing relevant information. Such a procedure is equivalent to the
transformation of data from an unstructured representation to a
structured network representation (nodes and edges).\\
\indent Thus, the task is to encode as much information of interest as possible,
in this case the existence of structure over a random distribution of
galaxies\\
\indent There is no universal technique to construct a network for this
kind of data, however the major steps to consider are the following: (\textit{i})
Capture similarity between data points; (\textit{ii}) Adopt some rules
based on a similarity function for establishing the links between data
points; (\textit{iii}) Implement some criteria to judge
whether the network is properly built, analogous to a
``goodness-of-fit'' procedure for approximation.\\
\begin{figure}

\includegraphics[width=0.5\textwidth]{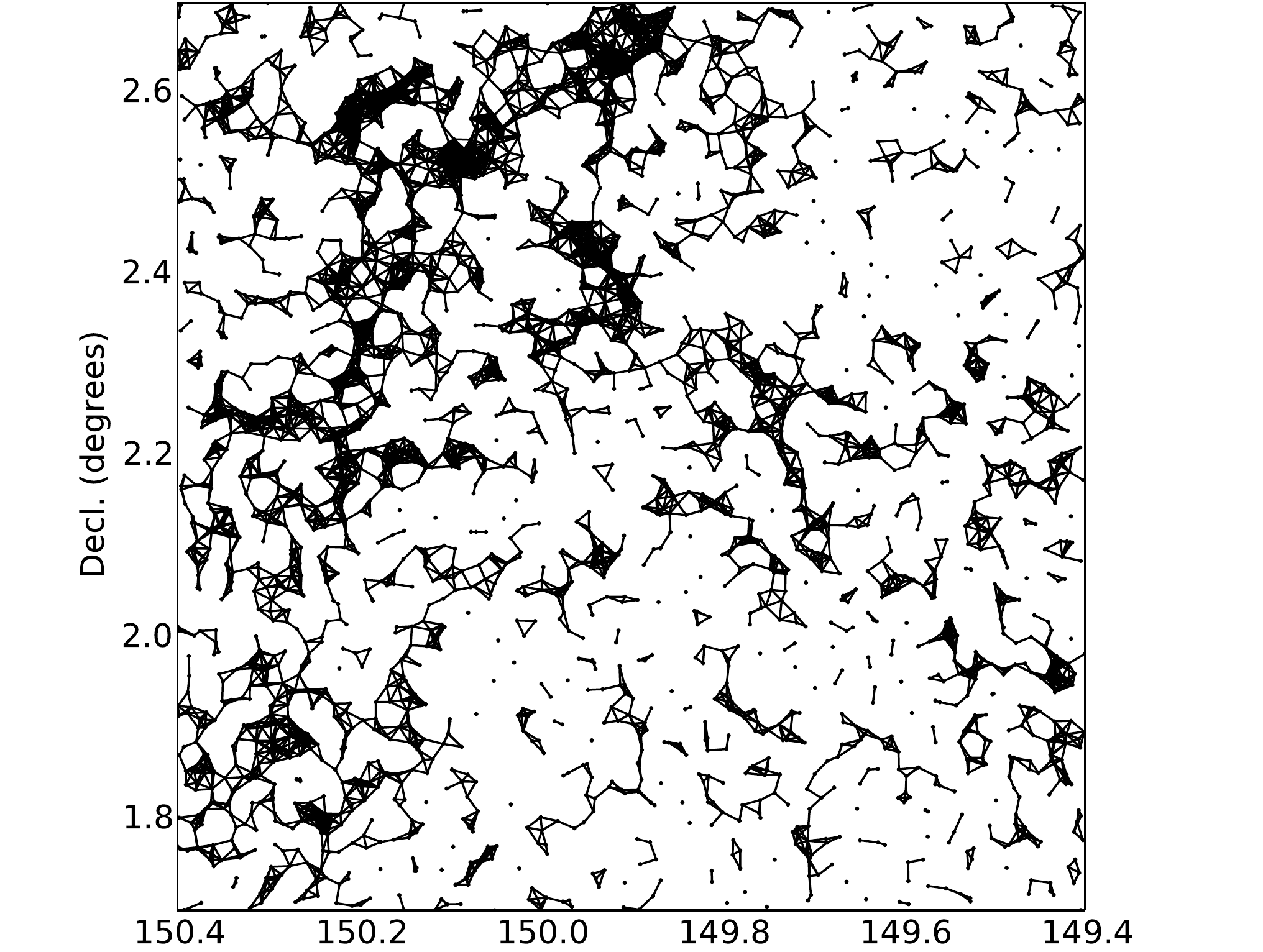}
\includegraphics[width=0.5\textwidth]{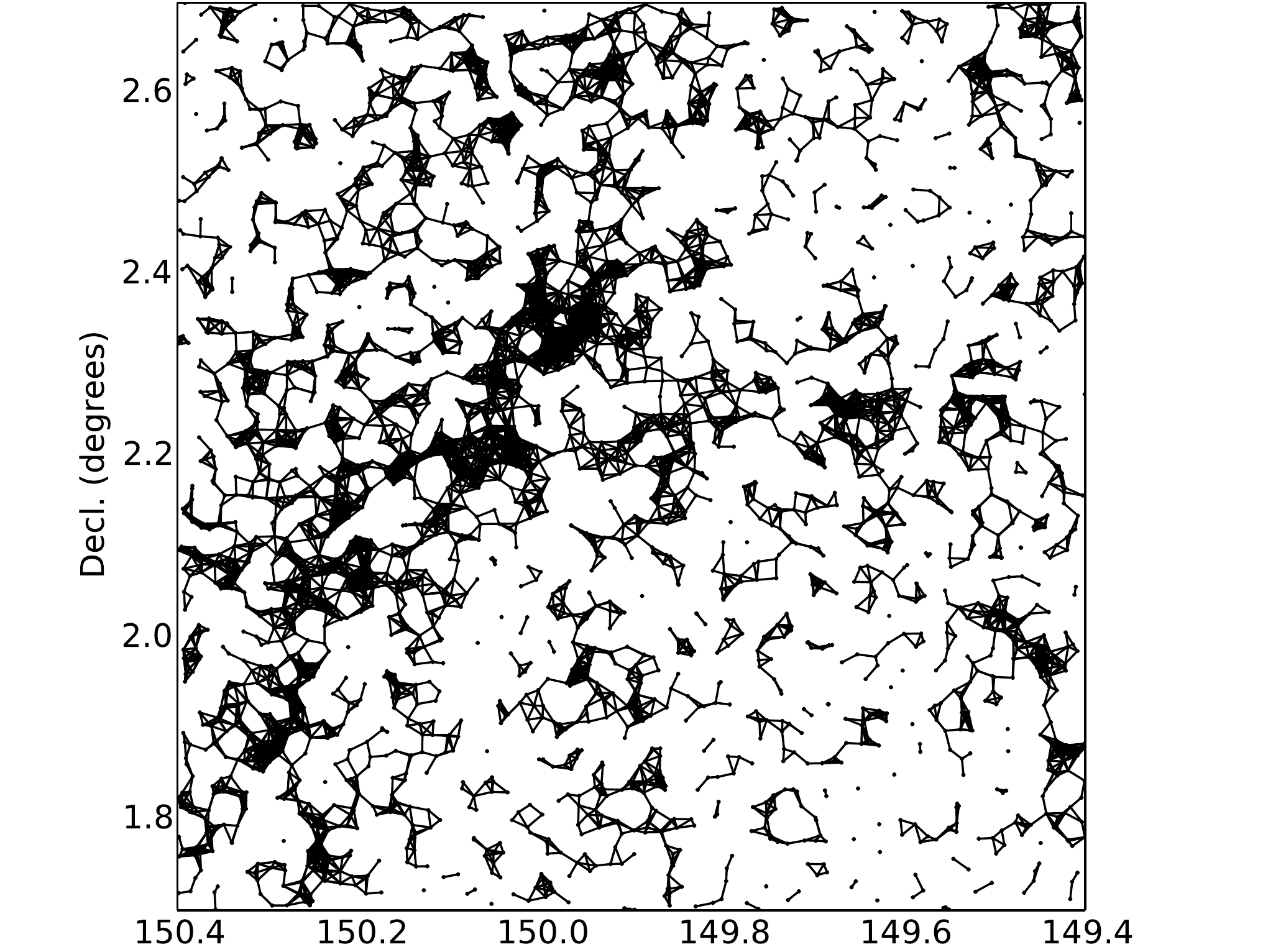}
\hspace*{-0.2cm}
\includegraphics[width=0.52\textwidth]{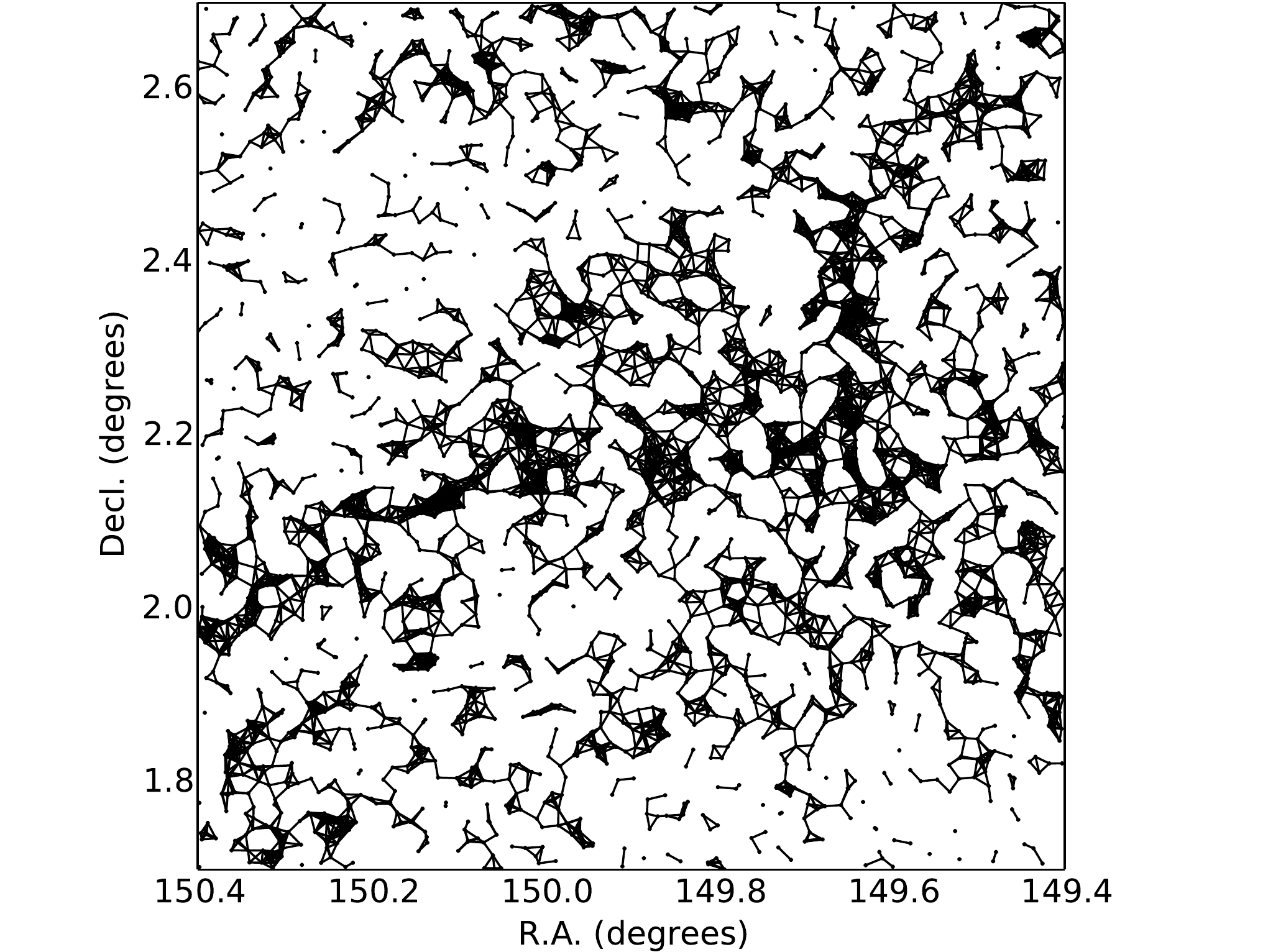}
\caption{Complex networks constructed on the base of the redshift
  slices $0.88 \le z_1 < 0.91$, $0.91 \le z_2 \le 0.94$, and
  $0.94 < z_3 \le 0.97$ (from top to bottom) from the COSMOS field
  using linking length of $0\overset{^\circ}{.}0216$. The middle
  figure recovers network that formerly obtained in
  \protect  \cite{Hong15}. } \label{fig1}
\end{figure}
\indent Different techniques for constructing complex networks from a galaxy
survey are discussed in \cite{Coutinho2016} on the basis of Illustris
cosmological simulation, and it was shown that proximity is the most
relevant similarity criterion for galaxy property studies. So, in this
paper we apply a similarity parameter of proximity, called
``linking-length''.\\
\indent Here, an undirected network is constructed by generating edges between
nodes, if and only if, the Euclidean distance between two nodes is
less or equal to the prescribed linking-length, which is fixed. This
simple recipe for analysing clustering was used for decades as
``top-hat filtering'' \citep{Bardeen1986} and is closely related to
the ``friend-of-friend'' algorithm \citep{Press1982}, used for the
study of large-scale structures. In the context of unsupervised
machine learning, the same approach is applied in density-based data
clustering algorithms, like DBSCAN or OPTICS \citep{dbscan, optics} as an
$\varepsilon$-radius method.\\ 
\indent So, hereafter a fixed linking length is predefined to be equal to
$0\overset{^\circ}{.}0216$, this corresponds to a linear scale of 1.2
Mpc in the standard $\Lambda$CDM cosmology.  This value was derived in
\cite{Hong15} from particular Poissonian distribution of node degree,
the closest to the observed one in the dataset.\\ 
\indent Such a method is proven to be robust to noise, albeit it is claimed in
\cite{Hong15} to not be universal, i.e. different samples would
require different linking length. Here, we implement a goodness-of-fit
measure based on the detection of a large connected cluster, or
``giant component'', which is an indication of structure in the
network. Also, the network should not be over-connected, or in other
words it should be as sparse as possible in order to accurately
reflect the relations between the nodes, and moreover be robust to
noise.\\
\indent Here, we investigate three cosmic networks constructed for the
different redshift slices using the same linking length calculated for
central slice (see also section 4.4). This allows to have consistency between samples and
enables comparison and tracking differences across samples. As the
outcome does not depend critically on precise value of linking length
and redshift slices are adjacent such simplification does not
introduce bias. In Fig.\ref{fig1} we show the cosmic networks
generated using this prescribed linking length $l$ for each redshift
slice. In the remainder of this section we will introduce the main
metrics used in network science to quantify different features.

\subsection{Network Metrics}

One of the remarkable features of most complex networks is their
heterogeneity. It leads to many unusual properties within
networks. Below, we will introduce some characteristics that will be
used to quantify network properties.

\subsubsection{Size}

The network is globally described by the numbers of nodes and edges,
$n$ and $m$ respectively. For the connected part of a network, 
there is always a path between any pair of nodes $i$ and $j$. The
shortest path length $\ell_{ij}$ between two nodes $i$ and $j$ can
then be described as the shortest route in terms of number of steps to
go from $i$ to $j$. The average path length $\langle \ell \rangle$ is
then the average number of steps along the shortest path for all possible
pairs of nodes belonging to the connected part of the network. It
gives a measure of how closely related nodes are to each other. Below,
we will calculate path lengths along the largest connected cluster of
the network (Giant Connected Component, GCC).

The equation used to
compute this quantity is:
\begin{equation}\label{2}
  \langle \ell \rangle = \frac{2}{g(g-1)}\sum_{i \neq j \epsilon GCC} \ell_{ij},
\end{equation}
where $g$ is the total number of nodes in GCC, $\ell_{ij}$ is the
shortest path between nodes $i$ and $j$, and the summation is performed
over all nodes belonging to the GCC.

This can then be compared with the average path length for a classical Erd\"os-R\'enyi random network $\langle \ell_r \rangle$ \citep{Erdos59} of the
same size, where links are randomly assigned between nodes.
\cite{Fronczak} have found it to be:
\begin{equation}\label{3}
  \langle \ell_r \rangle = \frac{\ln(g)-\alpha}{\ln(\langle k \rangle) + 0.5},
\end{equation}
where $\alpha \approx0.5772$ is the Euler-Mascherroni constant
\citep{Weisstien2002} and $\langle k \rangle$ is the mean node degree
defined in 3.2.2.

Another quantity that can be used to characterise the extent of a
network is the longest shortest path between any two nodes, sometimes
called the diameter of network, $D$. This path may provide an elegant description of the ``back-bone'' of the largest cluster in the cosmic network. In Table
\ref{tab} we list the quantities $n$, $m$, $\langle \ell \rangle$,
$\langle \ell_r \rangle$, $D$ and $g$ determined for all redshift
slices. Percentages in brackets next to values of $D$ and $g$ indicate
the portion of nodes belonging to the GCC.

\subsubsection{Centralities}

The importance of different nodes in a network can be determined by
their centralities. For one of the COSMOS galaxy samples, $z_2$, the centralities of nodes have already been
considered by \cite{Hong15}. Here, besides calculating centralities
for two more neighbouring redshift slices we evaluate not only
their point estimates but also assess their distributions. In turn,
this will allow us to compare galaxy samples in order to investigate
how these metrics differ in other redshift slices.  The centralities
we consider are Degree, Betweenness and Closeness, see
\cite{Brandes2001} and definitions below.

The degree centrality provides information on the connectivity of a
network within a localised area:
\begin{equation}
C_d(j) = \frac{k_j}{n-1},
\end{equation}
where $n$ is the number of nodes in the network and $k_j$ is the
degree (number of links adjacent) of node $j$,  determined in terms of an adjacency matrix
$\hat{A}$ as follows:
\begin{equation}\label{3.1}
k_j=\sum_{i} A_{ij}\, ,
\end{equation}
here and below, when not explicitly specified, the summations indices
span the entire network. For a network of $n$ nodes, $\hat{A}$ is an
$n \times n$ matrix with elements $A_{ij}=1$ if there is a link
between nodes $i$ and $j$ and $A_{ij}=0$ otherwise.  Table \ref{tab}
gives the mean values $\langle k \rangle$, $C_d$ and their standard
deviations and standard errors (in brackets) for each network.

The betweenness centrality defines how important a node is in terms of
connecting other nodes via shortest path lengths:
\begin{equation}
C_b(j) = \sum_{s,t(s \neq t \neq j)}\frac{\sigma_{st}(j)}{\sigma_{st}},
\end{equation}
where $\sigma_{st}$ is the number of shortest paths between nodes $s$
and $t$ and $\sigma_{st}(j)$ is the number of shortest paths between
nodes $s$ and $t$ that go through $j$.

The closeness centrality reveals how central a node is in the
network. Within any sub-connected component $\mathcal{F}$ of $f$ nodes
it is defined as:
\begin{equation}
  C_c(j) = \frac{f-1}{n-1}\frac{f-1}{\sum_{t \epsilon \mathcal{F}}\ell_{jt}},
  \label{CC}
\end{equation} 
If the network is disconnected, as is the case for our networks, the first
term will act to normalise the centralities for each fully connected
subcomponent.

\subsubsection{Correlations}

Correlations within networks can be investigated using different
techniques, implying both global and local characteristics. The Clustering coefficient of a network, in comparison with it random counterpart, can aid in quantifying the existence of structure within the local vicinity of a given galaxy and thus estimate its topological environment.

Local correlation is estimated by determining the clustering
coefficient of an individual node:
\begin{equation}
  \label{eq:CC}
  C(j) =\frac{2y_j}{k_j(k_j-1)} ,
\end{equation}
where $k_j\ge 2$ is degree of node $j$ and $y_j$ is the number of
links between neighbouring nodes of node $j$. When $k_j < 2$, then
$C_j = 0$ by definition. Averaging over all nodes in the network
yields a mean clustering coefficient for the whole network,
$C = \frac{1}{n}\sum_{i=1}^{n} C(i)$, the global characteristic of the
network.

\begin{table*}
  \centering
  \caption{ Network metrics for three galaxy samples at different
    redshift, $z_1$, $z_2$, and $z_3$, along with mean values for
    colour index and stellar masses.  Here, $n$ is number of nodes,
    $m$ is number of edges; $\langle \ell \rangle$,
    $\langle \ell_r \rangle$ are mean shortest path of GCC for real
    and random networks accordingly; $g$ is number of nodes and $D$ is
    diameter (maximal shortest path length) in the GCC; $k$ is node
    degree; $C_d$, $C_b$ are the degree and betweenness centralities;
    $C_{c1}$, $C_{c2}$ are closeness centralities for the distribution
    of fragmented clusters and GCC; $C$, $C_r$ are the mean clustering
    coefficients for real and random networks accordingly; $r$ is
    assortativity. The $Colour_{1}$ and $Colour_{2}$ are mean colour
    indexes for both modes of bimodal distribution, as shown in
    Fig. \ref{fig66}; $\log\,M_{stellar}$ is the logarithm of mean
    stellar mass (in units of solar one).  Where it is applicable, in
    brackets the standard deviation ($\sigma$) and standard error (SE), or
    percentages to indicate the portion of nodes involved in
    particular component are given.}
 \label{tab}
 \begin{tabular}{|l|l|l|l|l|}
   \hline
   \hline
   & 0.88$\le z<$0.91 &  0.91$\le z\le$0.94 & 0.94$< z\le$0.97 \\
   & Mean \quad [$1\sigma$, SE]& Mean \quad [$1\sigma$, SE]& Mean \quad [$1\sigma$, SE]\\
   \hline
   \hline
   $n$ & 3318 \quad  & 3678 \quad  & 3606 \quad  \\[2pt]
   $m$ & 11747 \quad \tab  & 14317  \quad \tab & 12206 \quad \tab  \\[2pt] 
   $\langle \ell \rangle$ & 37.53 \quad & 33.6 \quad  & 39.87 \quad  \\[2pt]
   $\langle \ell_r \rangle$ & 3.06 & 3.00 & 3.12 \\[2pt]
   $g$ & 2079 \quad [63\%] & 2369 \quad [64\%] & 2828 \quad [78\%]  \\[2pt]
   $D$ & 116 \quad [3.5\%]  & 113 \quad [3.1\%]  & 117 \quad [3\%] \\[2pt]
   $k$ & 7.08 \quad [5.02, 0.087] & 7.79 \quad [5.68, 0.093] & 6.77 \quad [4.36, 0.071] \\
   \hline
   $C_d$ & 0.0021 \quad [0.0015, 0.00003] & 0.0021 \quad [0.0015, 0.00003] & 0.0019 \quad [0.0012, 0.00002] \\[2pt]
   $C_b$ & 0.0045 \quad [0.014, 0.00023] & 0.0037 \quad [0.0097, 0.00016] & 0.0066 \quad [0.016, 0.00026]\\[2pt]
   $C_{c1}$ & 0.0019 \quad [0.00012, 0.000033] & 0.0028 \quad [0.0018, 0.000029] & 0.0018 \quad [0.0013, 0.000047]\\[2pt]
   $C_{c2}$ & 0.018 \quad [0.0041, 0.000090] & 0.021 \quad [0.0052, 0.000086] & 0.021 \quad [0.0052, 0.000097] \\
   \hline
   $C$ & 0.604 \quad [0.263, 0.0048] & 0.612 \quad [0.261, 0.0043] & 0.603 \quad [0.264, 0.0044] \\[2pt]
   $C_r$  & 0.0021 & 0.0021 & 0.0019 \\[2pt]
   $r$ & 0.85 \quad & 0.86 \quad & 0.80 \quad \\
   \hline
   $Colour_{1}$ & 0.64 \quad [0.66, 0.012] & 0.63 \quad [0.68, 0.012] & 0.61 \quad [0.67, 0.012]\\[2pt]
   $Colour_{2}$ & 4.02 \quad [0.54, 0.033] & 4.20 \quad [0.61, 0.032] & 4.13 \quad [0.66, 0.036]\\[2pt]
   $\log\,M_{stellar}$ & 9.29 \quad [0.67, 0.012] & 9.50 \quad [0.69, 0.011] & 9.44 \quad [0.66, 0.011]\\
   \hline
\end{tabular}
\end{table*}

To this end, to determine how strongly correlated a particular network
is, we can compare the $C$ with $C_r$, where $C_r$ is clustering coefficient for a Erd\"os-R\'enyi random network of the same size. Random networks are characterised by low values of $C_r$ and $\langle \ell_r \rangle$. So, if $C$ substantially exceeds $C_r$ this indicates that the network is highly correlated meaning that links in this network tend to be highly clustered together. The value of $C_r$
is calculated by simply considering $C_r= \langle k \rangle/n$.

Another useful estimator for node correlations is assortativity, $r$
which is usually used to investigate whether nodes of a similar degree
tend to be linked together. This is similar to the Pearson correlation
coefficient:
\begin{equation}\label{4}
r = \sum_{i,j}\frac{A_{ij}(k_i-E[k])(k_j-E[k])}{E[k^2-E[k]^2},
\end{equation}
where $A_{ij}$ is the adjacency matrix elements and $k_i$ and $k_j$ are the
degrees of node $i$ and $j$ respectively, $E[k]$ is the mean node
degree, $\langle k \rangle$, and $E[k^2]-E[k]^2$ is the mean variance
of the node degree.

Thus, with complex networks we can analyse the structure in a galaxy sample as a whole and in more detail. In particular, extend analysis beyond the local density and quantify short-range anisotropy of the distribution by clustering coefficient. Furthermore, we can compare different samples, retrieving important information which could not previously be revealed via existing methods. In the following section we apply network metrics to classify topological environments. Moreover, the application of complex networks to the Cosmic Web analysis places the research into a more general context of complex systems thus creating opportunities to search for analogies between different phenomena that occur in systems of interacting agents of various nature.

\section{Results and Discussion}
\label{IV}

Our results for different network metrics are listed in
Table.~\ref{tab}, the columns represent three networks visualised in
Fig.~\ref{fig1}. As it was already discussed at the end of Section
\ref{sect:NetworkConstruction}, we use the same linking length for all
redshift slices to enable a comparative analysis and reduce undue bias.

In Table.~\ref{tab} we can see that unique, yet similar
samples of galaxies, produce networks with resembling characteristics. This serves 
as confirmation that we have a robust network generation method which
generates a network with sufficient structure and relevant
information. Meanwhile, such comparisons also point to the unbiasedness
of the network generation method that exhibits sufficient sensitivity in detecting structure within galaxy distributions.

By comparing the average clustering coefficients $C$ and average path
lengths $\langle \ell \rangle$ with their random counterparts $C_r$
and $\langle \ell_r \rangle$, we can see that generated networks are
similarly, highly correlated networks, with evident regular structures
within their GCCs. The GCC is analogous to the super cluster in a
network and the diameter $D$ is analogous to the spine of the largest
cluster. From Table~\ref{tab} we see that all networks have slightly
different GCCs with similar spines. This would indicate a variance in
largest cluster size between networks with $z_3$ having the largest
cluster and $z_1$ the smallest. Present analysis does not exclude the
possibility that all three GCCs found for slices $z_1$, $z_2$ and
$z_3$ belong to a single extended structure.

\begin{figure*}
\includegraphics[width=0.99\textwidth]{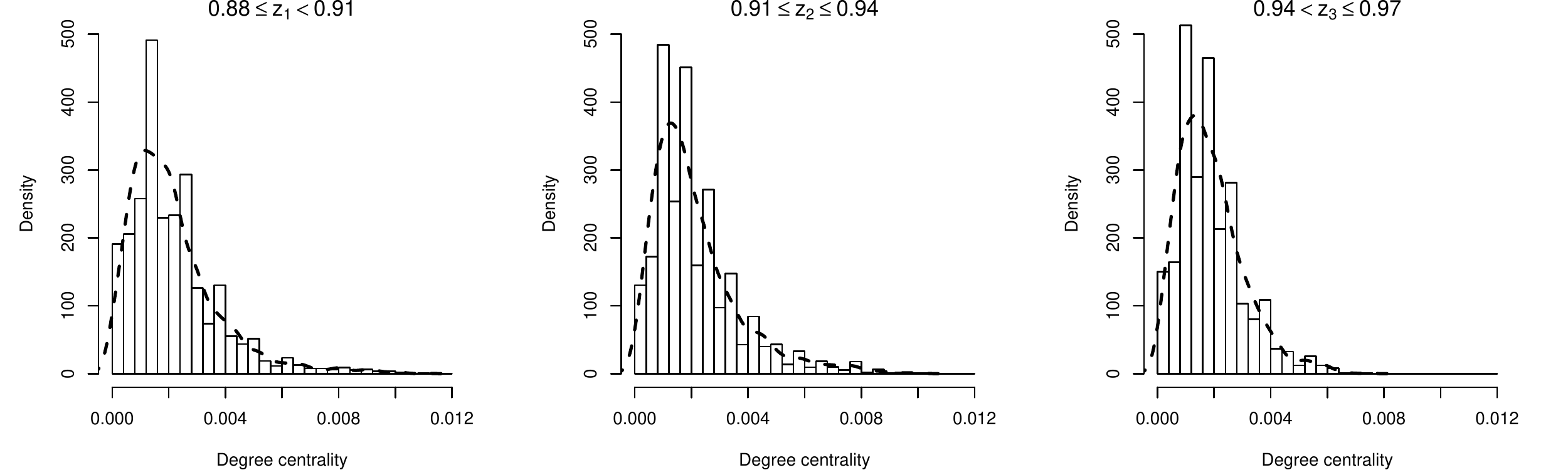}
\includegraphics[width=0.99\textwidth]{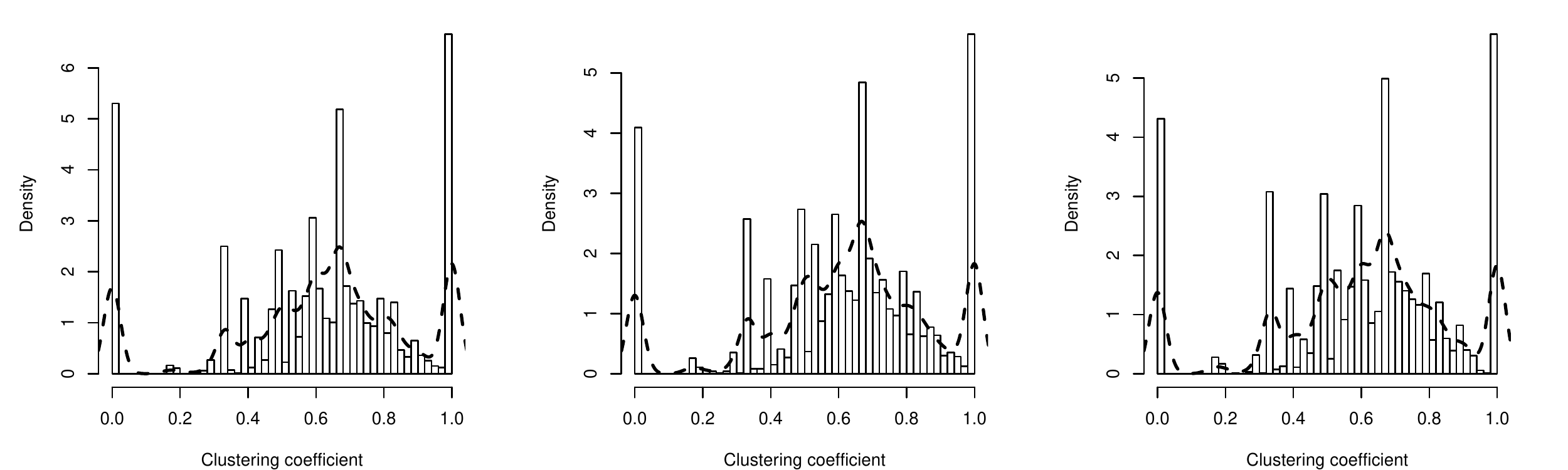}
\caption{The statistical distributions (histograms with density plots)
  of degree centrality $C_d$ and clustering coefficient for three
  ranges of $z$ from left to right: $0.88\le z_1 <0.91$,
  $0.91 \le z_2 \le 0.94$, and $0.94 < z_3 \le 0.97$.} \label{fig4.1}
\end{figure*}

We have computed the centrality measures for betweenness, closeness
and degree, which \cite{Hong15} consider in their paper, and estimated
their standard errors for three galaxy samples $z_1$, $z_2$ and
$z_3$. The distributions for centralities are summarised in Table
\ref{tab} and the distributions for degree centrality is shown in
Fig.\ref{fig4.1}.

\subsection{Degree centrality}

The degree centrality $C_d$ characterises the distribution of node degree
in a network, so the mean of such a distribution directly
relates to average degree $\langle k \rangle$. From Table~\ref{tab} we
can see that they are fairly similar with values of $7.08$, $7.79$ and
$6.77$ for increasing values of redshift. On inspection of
Fig.~\ref{fig4.1} (top row), the distributions on node centrality seem Poissonian in nature for all redshift slices, with $z_1$ and $z_2$ slices having more extended tails in comparison with $z_3$. This indicates that $z_1$ and $z_2$ have some really tightly packed galaxies within clusters whereas in $z_3$ the distances between galaxies are more evenly distributed within the clusters.

\subsection{Betweenness centrality}

\begin{figure}
\includegraphics[width=0.5\textwidth]{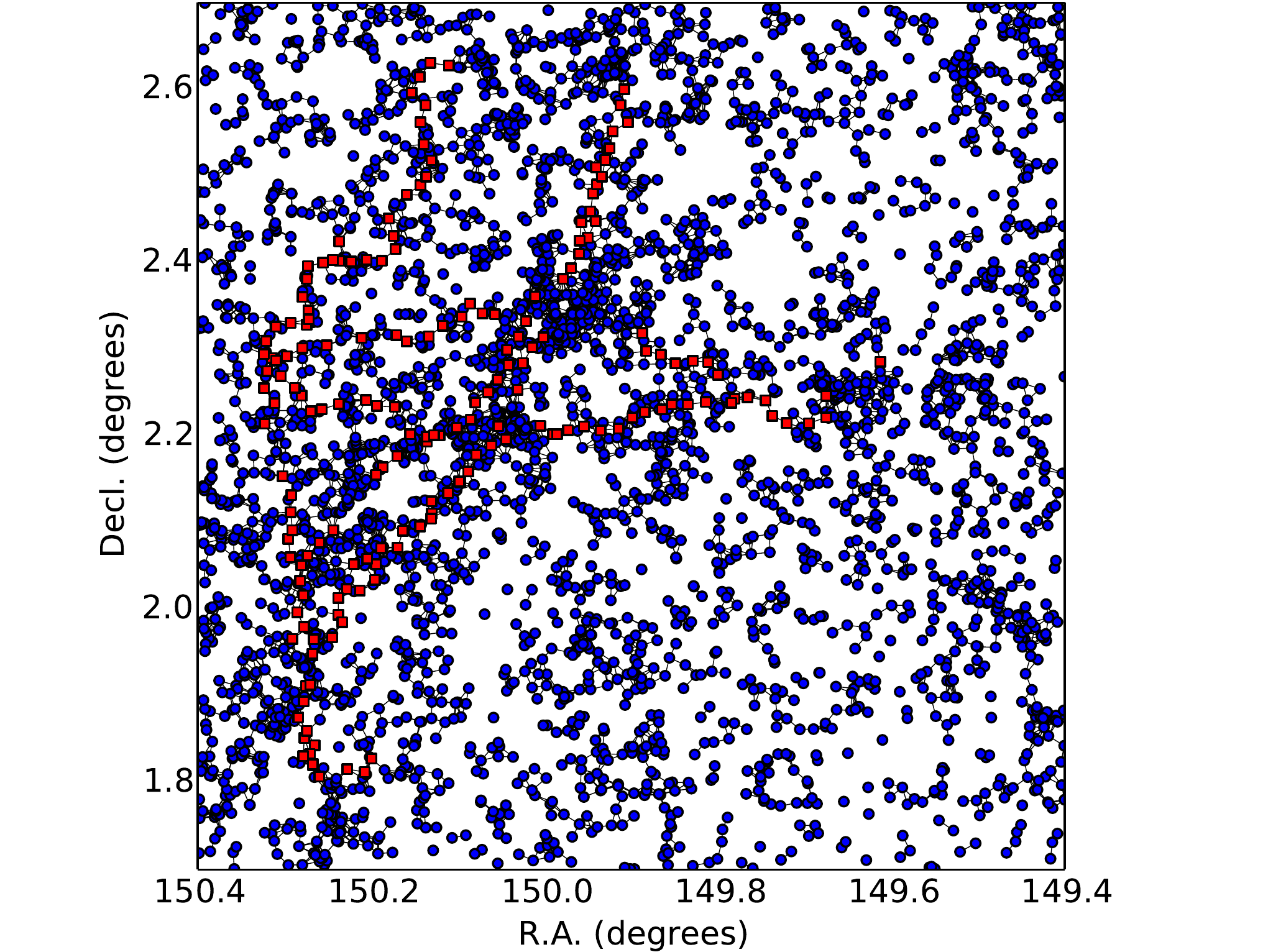}
\caption{Galaxies in $z_2$-slice with betweenness centrality greater
  than $0.02$ are red squares and galaxies with lower value are
  denoted by blue circles.}
\label{fig6}
\end{figure}

The betweenness centrality $C_b$ measures the importance of a node in
terms of maintaining connections between other nodes. In other words,
a node that is involved in a larger number of shortest paths will be
more important with respect to betweenness. Nodes which join two
large components/clusters together will also have a high betweenness
centrality. This is because many nodes exist in either of the two large 
clusters and hence many paths will have to traverse through these joining nodes.
This would not be the case if one of the clusters was small and the other large.
By this definition galaxies linking two larger clusters will display high betweenness
centrality.\\
\indent As it turns out from the analysis, the distribution
of the betweeness centrality is negatively skewed indicating a fewer number of high betweenness nodes. The galaxies with high betweeness might be
classified in astrophysical terms as filaments which join larger clusters
together. Fig.~\ref{fig6} depicts how galaxies with betweenness centrality greater then $0.002$ (shown by red squares) represent only a small portion of the galaxies and how they all tend to be galaxies that form paths between larger clusters.

\subsection{Closeness centrality}

We find that the distribution of closeness centrality $C_c$ is apparently bimodal
with two peaks centered about the values $C_{c1} \approx 0.002$ and
$C_{c2} \approx 0.02$. They are characterised by different
widths, leading in turn to different variance of the distributions
(see Table~\ref{tab}).

As it follows from a thorough analysis of the data, the
population of galaxies that belong to the second peak corresponds to
the largest connected component of the network (GCC). In turn, the
nodes in the centre of the GCC are characterised by shorter distances
to the rest of the nodes, leading by Eq.~($4$) to larger values of
$C_c$. The periphery nodes are characterised by larger distances to
the rest of the nodes, therefore they have smaller values of $C_c$.

In a similar way, one can identify the population of galaxies that
give rise to the first peak in the $C_c$ distribution. These are the
galaxies that belong to the smaller clusters, that are not attached to
the GCC. Here, the central nodes of the clusters correspond to the
right wing of the first peak and the periphery nodes are those
contributing to the left wing. The possibility to find two distinct
populations in the distribution is caused by the difference in sizes
of the GCC and that of the rest of the network. The larger the difference,
the more distinct the peaks. Indeed, as one can see from
Table~\ref{tab}, the largest size of GCC ($78\%$) is find for the redshift interval $z_3$. This which corresponds to the case where the gap between the two peaks is most pronounced.

\subsection{Clustering coefficient}

As it follows from Eq.~(\ref{eq:CC}), the clustering coefficient $C(j)$
counts the ratio of triangles of connected nodes to all possible
triples in a given cluster. In this way, the clustering coefficient is
a useful measure for the correlation on a {\em local} level or
\emph{short-range} correlation. It provides information on elementary
substructures (patterns) that appear in the network. In Table \ref{tab} observing clustering coefficient, one can see 
the presence of pervasive pattern-groups of tightly
connected galaxies on different sites (see also Fig.~\ref{fig1}) since
the high values of average clustering coefficient are obtained for all
redshifts slices.

\begin{figure}
\begin{center}
  \includegraphics[width=0.5\textwidth]{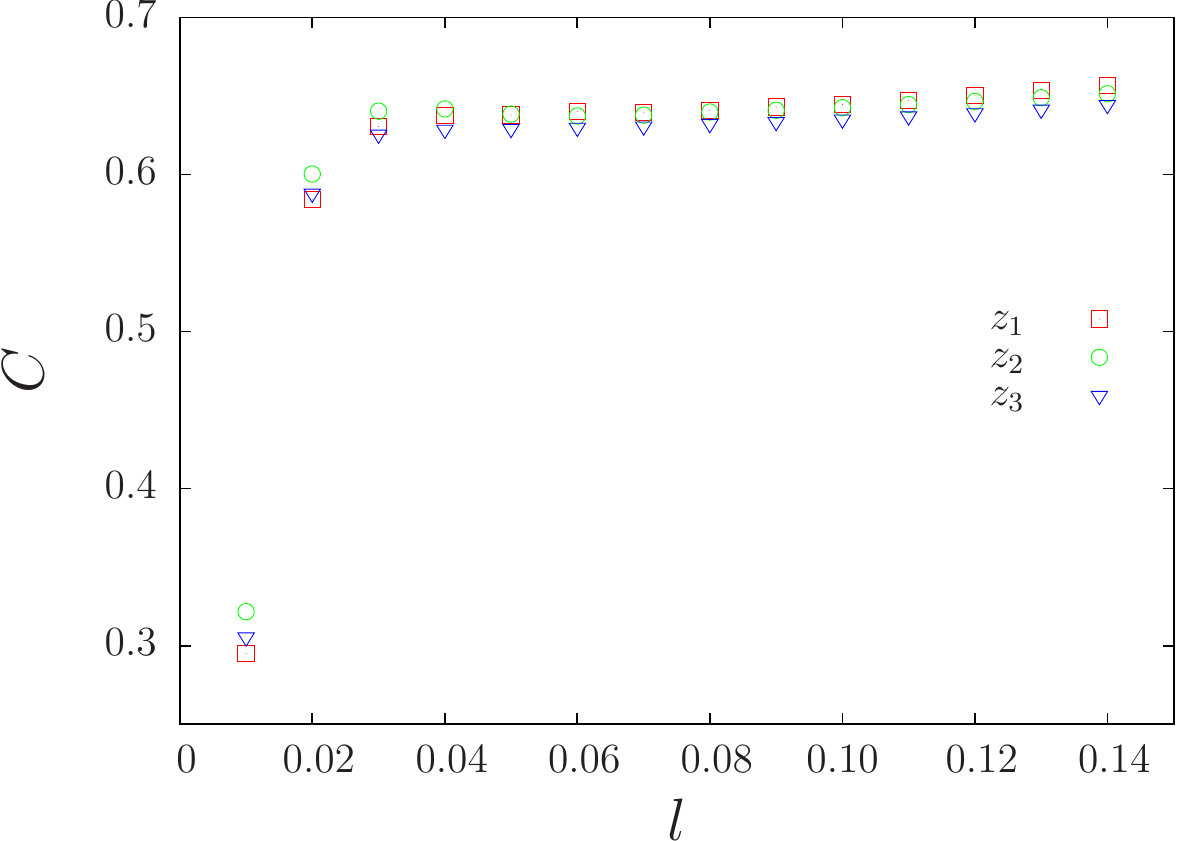}
  \caption{Clustering coefficient $C$ for all three redshift slices as
    a function of linking length $l$.}\label{fig3}
\end{center}
\end{figure}

Before continuing the discussion about the actual properties of $C$ for
the networks under consideration, let us return back to the origins of
network construction. As it was mentioned in
Section~\ref{sect:NetworkConstruction}, the choice of linking length
$l$ is crucial in defining the network topology, and it appears to be
particularly important with respect to correlation. Indeed, for a
small $l$ the network is just a set of disconnected nodes and
therefore $C=0$, while as it follows from Eq.~(\ref{eq:CC}) for large
$l$ one arrives at the complete graph where $C=1$.  In Fig.~\ref{fig3}
we illustrate this by plotting $C$ as a function of $l$ for all
redshift slices. One can see that $l=0\overset{^\circ}{.}0216$
chosen to construct Fig.~\ref{fig1} correspond to $C_1=0.604$,
$C_2=0.612$, $C_3=0.603$ at $z_1$, $z_2$ and $z_3$ accordingly.  So,
this value of $l$ appears to be optimal as shown first in
\cite{Hong15} and further supported by our analysis.

The local clustering of each node can also be considered in an effort
to help construct robust methods of defining substructures within the
cosmic network, or, for making selections to represent certain type of
environment.  Histograms for clustering coefficient in
Fig.~\ref{fig4.1} depict complex discrete distributions with three
main peaks at $0$, $0.66$ and $1$. In most cases galaxies with
clustering coefficient $C_i < 0.1$ have less then two neighbours, so
they are located in sparse environments, where mean distance between
galaxies is larger than the linking length.  This selection can be
called ``stand-alone'' galaxies represented by singlets and dumbbells
residing mostly in sparse regions. The nodes with clustering coefficient
ranging in $0.1\div0.9$ indicate galaxies that are intermediately
packed next to one another. Galaxies with a clustering coefficient
larger than $0.9$ tend to highlight small clusters, or in other words
participate in some ``cliques''. Thus, we make three selections of
galaxies, and analyse them below with regard to galaxy properties.

\begin{figure}
  \includegraphics[width=0.46\textwidth]{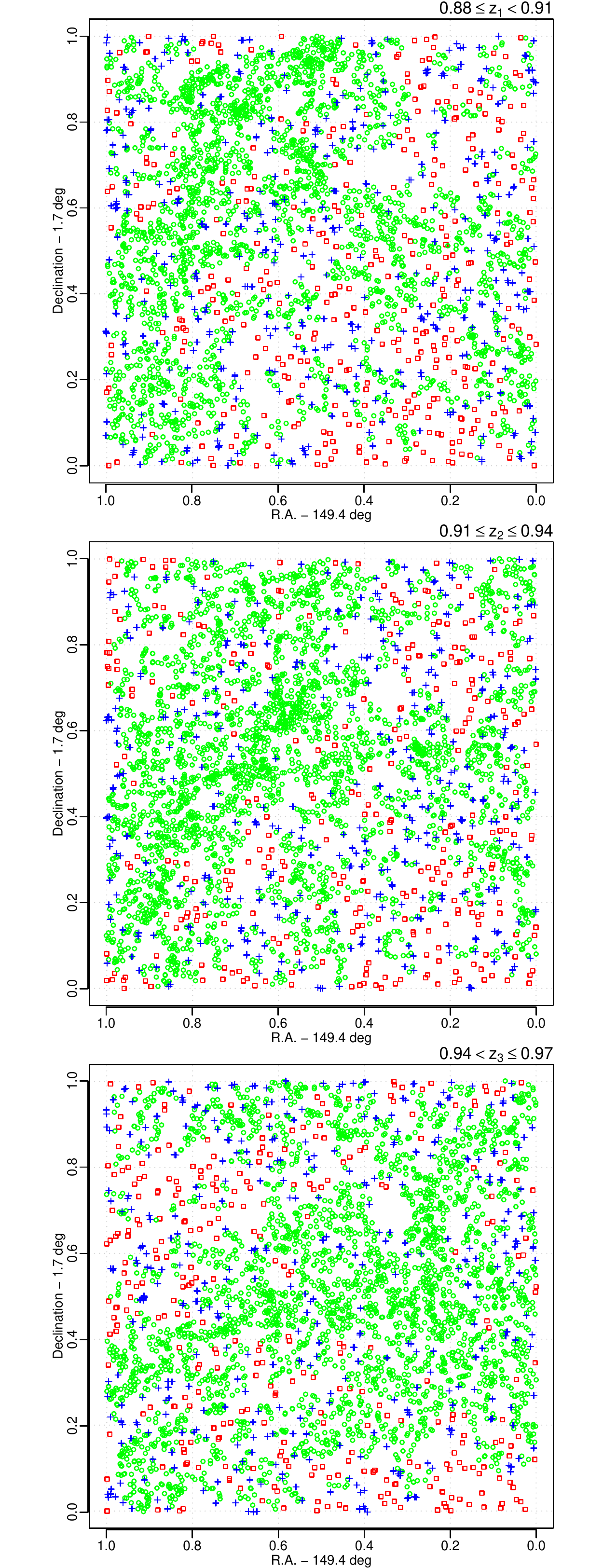}
  \caption{Galaxies from different selections marked according to
    their clustering coefficient. Red squares denote the
    ``stand-alone'' galaxies, green circles denote the galaxies with
    interim values of clustering coefficient, and small ``cliques''
    are denoted by blue crosses.}
\label{fig2}
\end{figure}

In Fig.~\ref{fig2} three selections of galaxies are mapped onto
spatial distributions, for each of three redshift slices. It is
noticeable, that nodes within denser clusters do not necessarily
exhibit higher clustering coefficient than their sparser
counterparts. The main reason is the fixed linking length. For
example, in these large clusters node $i$ will link to all nodes
within prescribed linking length including node $j$ on the edge of
linking length.  However, as the linking length is smaller than the
size of cluster, node $j$ will link to other nodes in this cluster
which are unreachable for node $i$ (not all neighbours of $j$ will be
linked to node $i$). Thus, counter intuitively, rather smaller
clustering coefficients are seen among highly clustered galaxies, and
clustering coefficient takes the highest values in smaller clusters at
the edges of voids.

\subsection{Average path length}

The evaluation of average path length makes sense only for the giant
connected component, because disconnected nodes will have no paths
between them, which mathematically leads to infinite lengths.
According to Table~\ref{tab}, $\langle \ell \rangle$ ranges between
$33$ and $40$ for different slices, to be compared with the
$\langle\ell_r \rangle$ of a random network of the same size.

In network theory significant amounts of attention have been paid to
the idea of small worldedness \citep{Watts98}: a network can be both
highly correlated on a local level (i.e. nearest neighbour level) and
exhibit relatively small $\langle \ell \rangle$ at the same time.
When $C$ of a network exceeds randomly expected, $C \gg C_r$ and
$\langle \ell \rangle$ is close or smaller than randomly expected,
$\langle \ell\rangle \lesssim \langle \ell_r \rangle$, then a network
is said to be small world in nature.

The cosmic networks do not display small world characteristics. All
three networks satisfy the first condition of small worldedness in
that they are far more correlated then randomly expected. However,
these networks fail on the second condition in that
$\langle \ell \rangle$ are all much larger than randomly expected and
so can not be considered to be small world in nature. Therefore, the
cosmic network is a large world in this context. This could well be a
result of the constraint that is imposed by linking length, as this
does restrict galaxies outside a certain distance from being linked
and could be a contributing factor in why the network is a large
world.

\subsection{Assortativity}

\begin{figure}
\centering
\includegraphics[width=0.46\textwidth]{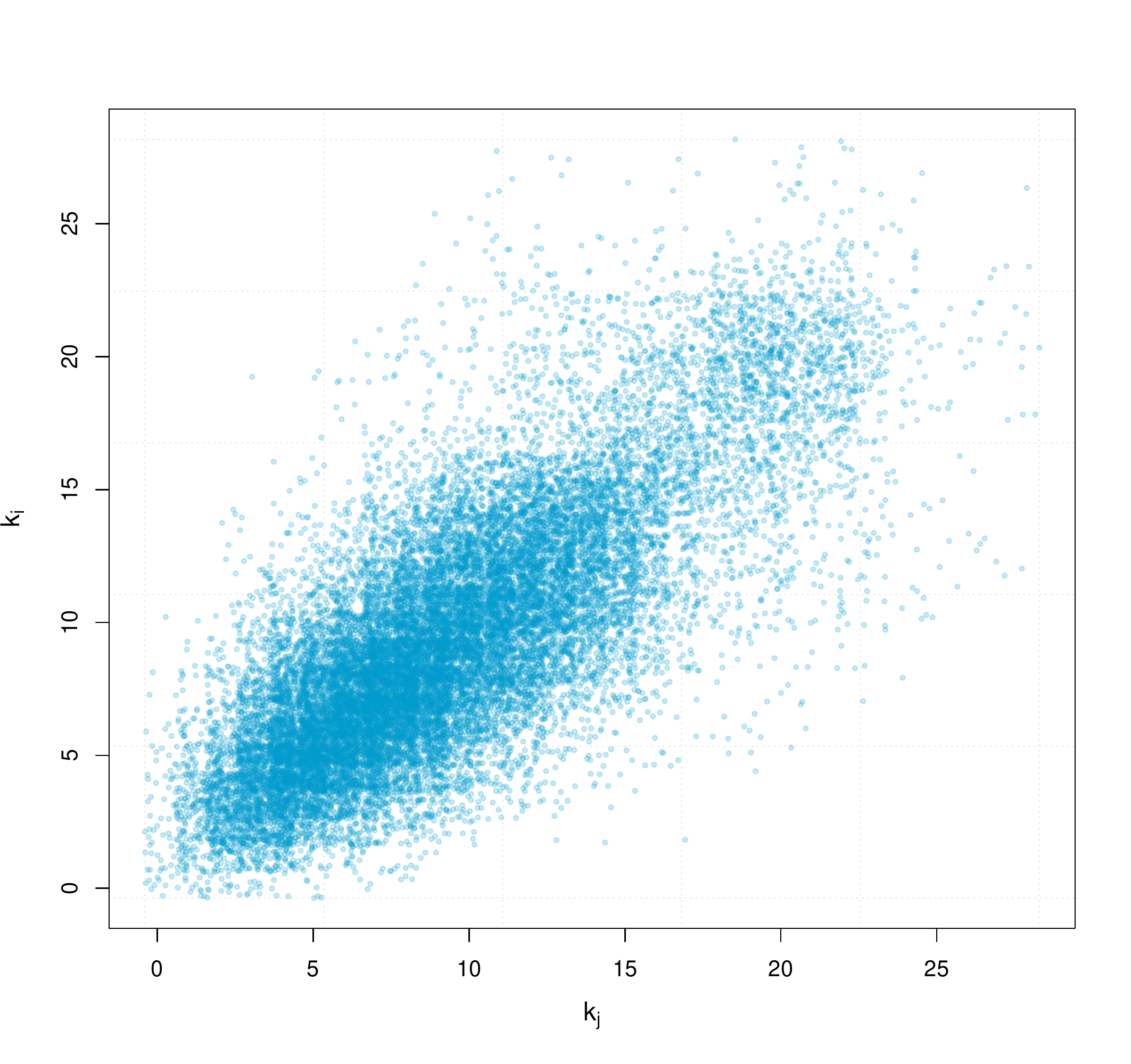}
\caption{Scatter plot for degrees of connected nodes: $k_i$ and $k_j$ for the $z_2$ redshift slice. \label{fig4}}
\end{figure}

For a disassortative network the value of $r$, Eq.~(8), is
negative indicating that nodes of low degree tend to associate with
nodes of high degree. In turn when this value is positive this
indicates an assortative network where nodes of similar degree link
with one another. Fig.~\ref{fig4} provides a qualitative perspective
where it can be clearly seen that the cosmic network displays positive
correlation and this can be further confirmed quantitatively in
Table~\ref{tab} with $r$ for all redshifts being $\geq 0.80$. This
indicates that in the cosmic network galaxies with a similar number of
links tend to be connected to one another.

\subsection{Astrophysical quantities vs topology}

\begin{figure*}
  \includegraphics[width=0.99\textwidth]{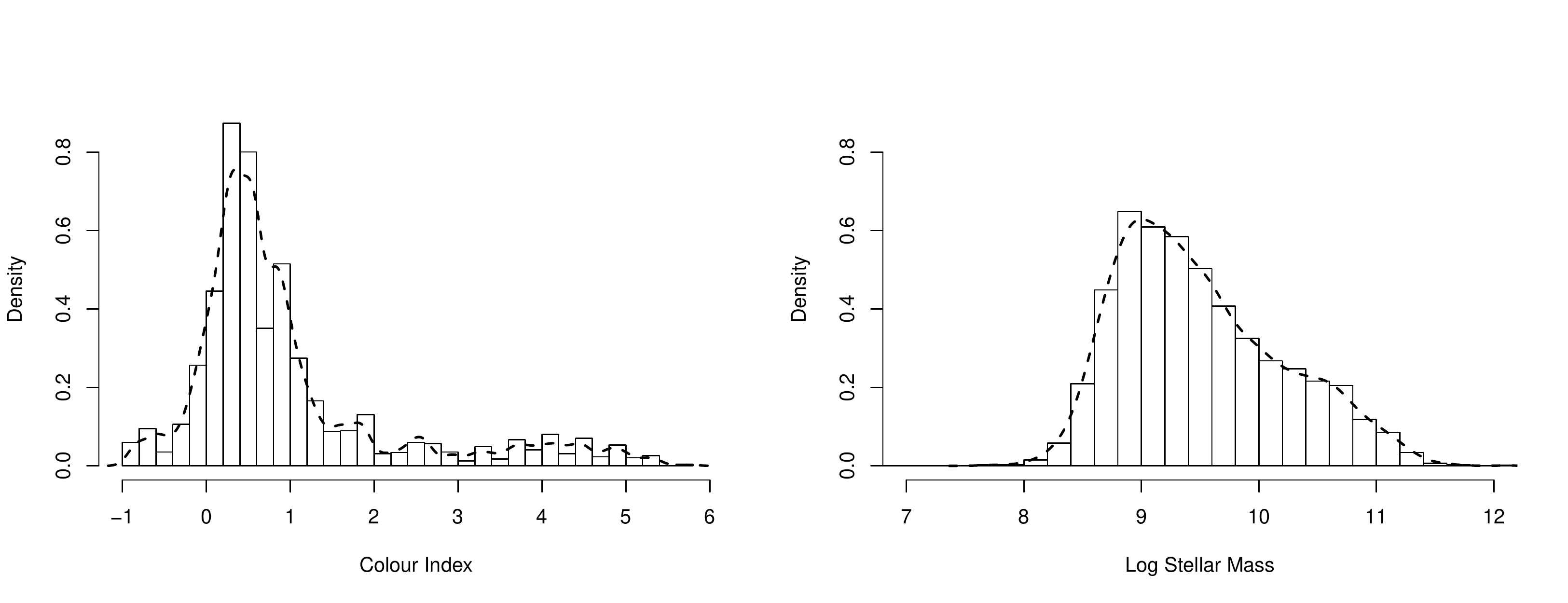}
  \caption{The statistical distributions of galaxy parameters: colour
    index and stellar mass for redshift slice
    $0.91 \le z_2 \le 0.94$.} \label{fig66}
\end{figure*}

Another goal of this research is to investigate how galaxy
properties (hereafter the stellar mass and colour index) relate to the
topological environment of galaxies (hereafter topological refers to
selections according to clustering coefficient). The relationship between
galaxy properties and network centrality measures have been considered
by \cite{Hong15} for the $z_2$ slice. Here we take a different approach,
based on clustering coefficient, and apply it to three samples of
galaxies.

Before embarking into the analysis, we need to address a number of
the limitations caused by the nature of the data. The exploration of
clustering coefficient (bottom panel of Fig.~\ref{fig4.1}) reveals its
discrete and highly non-uniform distribution, meanwhile the
astrophysical parameters are continuous variables with non-trivial
distributions (especially colour index, see Fig.~\ref{fig66}). Given
that parametrical methods for multivariate analysis
e.g. correlation analysis, are definitely inapplicable, and even
though the application of non-parametrical methods cannot ensure feasible
results we are left these methods to apply.

Of course, we can seek for trends by analysing general differences
between distributions in samples, for instance by comparing their
means and standard deviations, as in Table~\ref{tab}. However, the
statistical significance of such differences is unknown.

The distributions of variables can be compared by means of
non-parametric methods based on empirical distribution function
(two-sample tests). At some confidence level, null hypothesis
significance testings estimate $p$-values to be used for rejecting the
null hypothesis, in this case that both selections are sampled from
the same population. Note, that such tests result in binary answers
(yes/no), seek to reject the null hypothesis, and should be taken
with a grain of salt since they assume the univariate nature of
variables.

Usually the Kolmogorov-Smirnov test \citep{Kolmogorov33,Smirnov48} is
used as a non-parametric test, as in \cite{Hong15}. Although this test
is universal tool, it has a number of limitations, and should be
cross-validated by other approaches, like Anderson-Darling
\citep{Anderson54} or Mann-Whitney-Wilcoxon \citep{Mann47,Wilcoxon45}
tests. 

\subsubsection{Distributions of galaxy parameters} 

We first analyse distributions for colour index and stellar mass
(Fig~\ref{fig66}), the means and standard deviations are included in
Table~\ref{tab}. The Hartigans' dip test \citep{Hartigan85} proves
that bimodalities in the colour distributions are statistically
significant: the null hypothesis of unimodality is rejected with
$p$-value $< 2.2\cdot10^{-16}$. The different heights of the peaks in
histogram imply heterogeneity of the data set, which may be drawn from
different populations.

With respect to colour index, non-parametric tests consistently
indicate the following: the hypothesis of a common distribution is
strongly rejected when comparing $z_1$ and $z_3$ samples, mildly
rejected for $z_1$ and $z_2$ samples, and mildly accepted for $z_2$
and $z_3$ samples. Therefore, the tests have revealed a weak but still
significant evolutionary trend for colour index over redshifts span.

Although the shape of distribution for stellar mass is simpler,
two-sample tests for stellar mass detect significant distinctions
over redshifts for all pair-wise comparisons except in the case of $z_2$ vs
$z_3$. Note that colour index is derived directly from observed
photometric measurements. Meanwhile, the stellar mass of galaxies is
computed from the same photometric data using approximations and
elaborate modeling of spectral energy distributions (SED).

\begin{table}
  \centering
  \caption{The results of Anderson-Darling tests ($p$-values) for
    colour and stellar mass distributions for different clustering
    coefficients selections: I $C=0$; II $0 < C < 1$; III
    $C=1$. The critical $p$-value equals 0.05.}
  \label{tab-KS}
 \begin{tabular}{c|c|c|c|c|c|c|}
   \hline
   &  \multicolumn{3}{c}{Colour} & \multicolumn{3}{c}{Stellar Mass} \\
   \hline
   &   $z_1$   & $z_2$     & $z_3$      & $z_1$      & $z_2$      & $z_3$ \\ 
   \hline
   I vs II    &  0.062  &  0.018  &  0.31   &  5$\cdot10^{-6}$ &  0.0048  &  0.00023  \\
   II vs  III &  0.29   &  0.025  &  0.37   &  0.0032  & 0.0018   &  0.014  \\
   I vs III   &  0.74   &  0.79   &  0.49   &  0.19    & 0.91     &  0.18  \\
   \hline
\end{tabular}
\end{table}

\subsubsection{Selections by clustering coefficient}
\begin{figure*}
\includegraphics[width=0.8\textwidth]{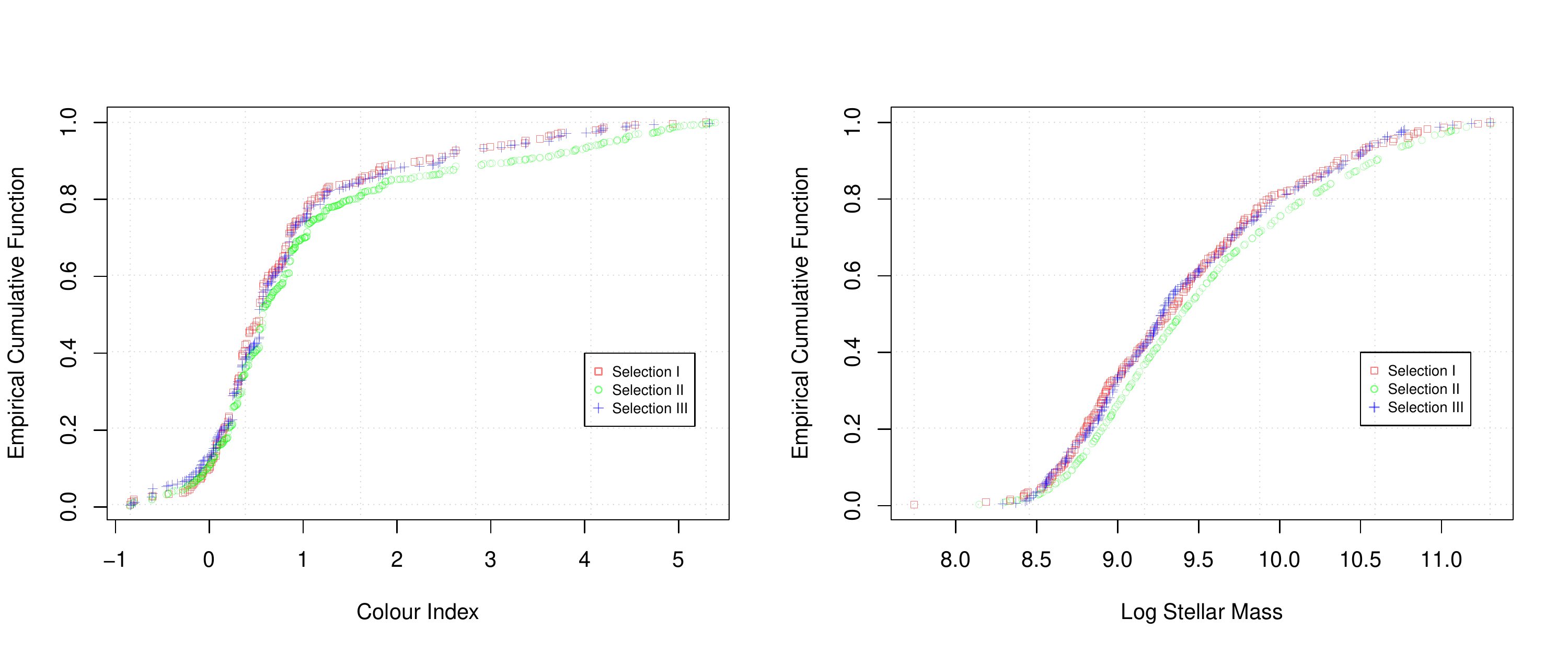}
\caption{The empirical cumulative distribution functions 
for colour index (left) and stellar mass (right) for different selections by local
clustering coefficient, designated by red squares, green circles and blue crosses for selections I, II and III respectively for the redshift sample $z_2$.}
\label{fig-KS}
\end{figure*}

Given the different nature of distributions we should follow two-step
a procedure in order to find out how colour index and stellar mass of a
galaxy are determined by clustering coefficient of the galaxy: split
the data set into three subsamples (or selections) according to local
clustering coefficient; then compare empirical distribution functions
of galaxy properties for different subsamples by two-sample tests.
Thus, each redshift slice was split into three subsamples: selection I
(stand-alone galaxies) $C=0$; selection II (intermediately packed
galaxies) $0 < C < 1$; selection III (compact cliques of galaxies)
$C=1$. Then distributions of different selections are tested for
equality in pair-wise manner.  Fig.~\ref{fig-KS} presents the
empirical cumulative distribution functions of colour index and
stellar mass for selections I (red squares), II (green circles) and
III (blue crosses) for redshift sample $z_2$.
\\
\indent
In Table~\ref{tab-KS} we present the results of the non-parametric
Anderson-Darling tests. Again, here the null hypothesis states that
subsamples are drawn from the same population, the alternative
hypothesis states the populations are different.  The $p$-value
indicates the statistical significance of test, if it is less than
$0.05$ the null hypothesis can be rejected with high degree of
confidence. Note that the magnitude of $p$-values do not reflect the
strength of the effect.
\\
\indent
We can deduce the following conclusions from Table~\ref{tab-KS}:
i) the samples of selections I and III (stand-alone and densely packed
in small groups of galaxies) are non-distinguishable, for all
$z$-slices, with respect to both colour and stellar mass; ii) the
distribution of stellar mass for selection II differs from selection I
and III across all $z$-slices; iii) the distribution for colour index
for selection II does not differ from selections I and III, expect when 
considering the $z_2$-slice.
\\
\indent
The weakness of the evolutionary effects is understandable since the
age differences of the nearest and farthest sample of galaxies do not
exceed 400 million years.  We have however to bear in mind the caution
expressed already at the beginning of the paper: the database used
here does not allow one to use coordinates of galaxies in 3D space
with high enough precision. Indeed, the 2D slices of the real-world
pictures (see Fig.~\ref{fig1}) result from the projection of their 3D counterparts. According to \cite{Scoville13}, the binning matched to accuracy of the redshifts, 
thus providing optimal signal-to-noise ratio. For the density estimation the 2D projections are linearly related to a 3D volume whereas for the topological environment that might not be the case. Despite of this obvious limitation 
one can still retrieve information on the correlations we are interested in.
\\
\indent

The research presented above allows one to approach probably the most important problem in cosmology, the mapping of the observable distribution of luminous matter to the underlying dark matter distribution, sometimes called the problem of biasing. The results here are derived from real-world observational data, so they are not just a description of the spatial structure, they encode information of extremely complex processes of star formation, gas and radiation transfer in different environments. So, our findings on the common behaviour in the evolution of stand-alone galaxies and cliques bring important confirmation for the Cosmic Web Detachment model \citep{Argon2016}, identifying the events of detachment in real observations.

\section{Conclusions}
\label{V}

Here we have analysed some observed part of the Cosmic Web (COSMOS
catalogue of galaxies \citep{Ilbert13}) by means of complex
network analysis. A major distinction of our study is that we analysed 
galaxy samples in the same region $1^\circ \times1^\circ$ of the
celestial sphere as the previous study of \cite{Hong15}, but for three
neighbouring redshift intervals $0.88\le z <0.91$, $0.91\le z\le0.94$
and $0.94 < z \le 0.97$, marked by $z_1$, $z_2$ and $z_3$ accordingly.\\
\indent We have developed and validated the robustness of our technique for
constructing complex networks from galaxy samples using a fixed linking
length method ($l = 0\overset{^\circ}{.}0216$). For each redshift slice we have calculated the
local complex network measures, namely degree, closeness and
betweenness centralities, clustering coefficient $C(j)$ as well as
the global measures, e.g.  average path length $\langle \ell \rangle$,
diameter $D$, average clustering coefficient $C$, number of nodes $g$
and diameter $D$ of the giant connected component GCC, mean node
degree $k$, assortativity $r$. \\
\indent We have not found firm evidence of evolutionary changes across
complex networks, either by comparing the distributions of the local
network measures or analysing global network measures. The main
reason maybe due to the insufficient differences in the
cosmological ages of galaxy samples.\\
\indent The comparison of the computed measures of our networks with
corresponding measures of random ones give us some global
characteristics of the Cosmic Web in the context of complex network
theory. Together these properties imply that constructed cosmic networks are not small worlds
in terms of network science but rather ``large worlds''. \\
\indent The size of Giant Connected Component (GCC) informs about the largest
cluster in a network, here it contains 63\%, 64\% and 78\% of galaxies
in $z_1$, $z_2$ and $z_3$ samples accordingly. The high value of
assortativity coefficient $r\sim0.80\div0.86$ means that in the cosmic
network galaxies with a similar number of links tend to be connected
to one another.  \\
\indent Most of the local network measures have non-Gaussian distributions,
often bi- or multi-modal ones (Fig.~\ref{fig4.1}). The local
clustering of each node $C(j)$ in the cosmic network shows a three
mode distribution which allows for the discrimination between singlets
and dumbbells of galaxies ($C=0$) on the one hand and cliques of
galaxies ($C=1$) on the other. So, the network metrics analysed here
allow for discrimination between topologically different structures.\\
\indent Another goal of our study was to analyse the impact of surroundings on
the astrophysical properties of galaxies, in particular colour indices and
stellar masses. Doing so, besides studying the obvious impact of the 
immediate neighbourhood of a galaxy (which can be and is done by means
of other methods too) we presented here an elaborated method to study
the subtle topological features of galaxy distribution beyond its local
density, as short-range clustering.\\
\indent The general analysis of trends in means and standard deviations of
colour indices and stellar masses across redshift slices $z_1$, $z_2$
and $z_3$ has not revealed substantial differences, see
Table~\ref{tab}. Meanwhile, the comparison of distributions via
non-parametric tests detects a weak evolutionary trend over the
redshift span 0.88$\div$0.97 for the colour index of galaxies.\\
\indent Comparison (with Anderson-Darling test) of the empirical distribution
functions for astrophysical characteristics by different selections
defined by the modes of clustering coefficient yields evidence of
consistent and statistically significant associations between astrophysical
quantities and topological selections, see Fig.~\ref{fig-KS} and Table
\ref{tab-KS}.\\
\indent In particular, it was shown that stand-alone galaxies with $C(j)=0$
(selection I) and galaxies densely packed in small cliques with
$C(j)=1$ (selection III) are not distinguishable by colour index and stellar mass distributions.\\
\indent Stellar mass distributions for galaxies with interim clustering
coefficient (selection II) differ from the corresponding distributions
in selections I and III. This difference holds for all redshift
slices. The analogous difference in colour index distributions holds
however only in the $z_2$ redshift slice. The latter $z_2$-sample has
been intensively studied by other methods in the papers
\cite{Scoville13} and \cite{Hong15}.\\
\indent The presented results demonstrate the promising use of complex network
theory in the study of the Cosmic Web. With the improving accuracy of redshift values for galaxies, we hope that in future, this will allow the cosmic network to
be studied in 3D which will in turn provide more accurate
results.

\section*{Acknowledgements}

This work was supported in part by the projects: 0116U001544 of the
Ministry of Education and Science of Ukraine (S.A. and B.N.); the FP7 EU IRSES project 612707 ``Dynamics of and in Complex Systems''
(R.dR., C.vF., and Yu.H.) and by the project DFFD 76/105-2017 "Complex network concepts in problems of quantum physics and cosmology". Authors thank the entire COSMOS
collaboration for available data at COSMOS Archive
http://irsa.ipac.caltech.edu/data/COSMOS.

\label{lastpage}

\end{document}